\documentclass{aa}
\usepackage{graphicx}
\usepackage{times}
\usepackage{mathptm}
\fontfamily{ptmcm}\selectfont
\begin{document}

\thesaurus{04.19.1; 13.18.2; 03.13.2}
	   
\title{The ATESP Radio Survey}

\subtitle{I. Survey Description, Observations and Data Reduction}

\author{I. Prandoni \inst{1,2}
	\and L. Gregorini \inst{3,2}
	\and P. Parma \inst{2}
	\and H.R. de Ruiter \inst{4,2}
	\and G. Vettolani \inst{2}
	\and M.H. Wieringa \inst{5}
	\and R.D. Ekers \inst{5}
        }
\offprints{I. Prandoni}
\mail{prandoni@ira.bo.cnr.it}

\institute{Dipartimento di Astronomia, Universit\`a di Bologna, via Ranzani 1,
I--40126, Bologna, Italy
\and Istituto di Radioastronomia, CNR, Via\,Gobetti 101, I--40129, 
Bologna, Italy
\and Dipartimento di Fisica, Universit\`a di Bologna, Via Irnerio 46,
I--40126, Bologna, Italy
\and Osservatorio Astronomico di Bologna, Via Ranzani 1, I--40126, 
Bologna, Italy
\and Australia Telescope National Facility, CSIRO, P.O. Box 76, Epping, 
NSW2121, Australia
}

\date{Received 07 March 2000 / Accepted 23 June 2000}

\titlerunning{The ATESP Radio Survey. I}
\authorrunning{I. Prandoni et al.}

\maketitle
\begin{abstract}

This paper is the first of a series reporting the results 
of the \emph{Australia Telescope ESO Slice Project} (ATESP) radio survey 
obtained at 1400 MHz with the \emph{Australia Telescope 
Compact Array} (ATCA) close to the South Galactic Pole (SGP) over the region 
covered by the \emph{ESO Slice Project} (ESP) galaxy redshift survey ($\delta 
\sim -40\degr$).  
The survey consists of 16 radio mosaics with $\sim 8\arcsec \times 14\arcsec$
resolution and uniform sensitivity ($1 \sigma$ noise 
level $\sim$79~$\mu$Jy) over an area of $\sim 26$ sq.~degrees.  
Here we present the design of the survey, we describe the mosaic observing 
technique which was used to obtain an optimal combination of uniform and high 
sensitivity over the whole area, and the data reduction: the problems 
encountered and the solutions adopted.   


\end{abstract}

\section{Introduction}\label{sec-intr}

Sky surveys play an important role in astronomy. Large collections of
objects allow reliable studies of the average properties 
of different constituents of the universe; new populations 
can be found and well defined
sub-samples can be extracted for further detailed analysis. \\
Surveys in the radio domain have yielded many important results in the
past, since the discovery of radio galaxies and quasars. 
The first samples of radio sources ($S\ga
1$ Jy) have demonstrated that classical radio galaxies are rare in the
local universe and strongly evolve with cosmic time both in density and 
luminosity (e.g. Longair \cite{Longair66}).
More recently, deep radio surveys ($S \la 1$ mJy) have shown that
normalized radio counts show a flattening below a few mJy,
corresponding to a steepening in the actual observed counts (see e.g. 
Windhorst et al. \cite{Windhorst90} for counts at 1.4 GHz).
This change of slope is generally interpreted as being
due to the presence of a new population of radio sources (the so-called
sub-mJy population) which
does not show up at higher flux densities (see e.g. Condon \cite{Condon89}).
To explain the new faint radio population several scenarios have been invoked:
strongly-evolving normal spirals (Condon \cite{Condon84}, \cite{Condon89}); 
actively star 
forming galaxies (e.g. Rowan-Robinson et al. \cite{Rowan93}); or a 
non-evolving population of local (z $<$ 0.1) low-luminosity galaxies 
(e.g. Wall et al. \cite{Wall86}).
The true nature of
the population is not well established. The same is true for the relative  
contributions from the above mentioned objects. Furthermore the source
space density inferred from the faint end of the local bivariate luminosity 
function for both spirals and 
ellipticals is not well known (see e.g. Condon \cite{Condon96}). 
Therefore it is not possible to estimate the local contribution to the counts  
(even if expected to be small) nor is there  a clear local reference frame 
for understanding evolutionary phenomena. \\
Unfortunately, due to the long observing times required to reach faint fluxes, 
the existing samples in the sub-mJy region are generally small. 
Table~\ref{rad-surv} shows a compilation of the largest 1.4 GHz surveys 
available in the mJy and
sub-mJy regime with the surveyed areas and limiting fluxes (note, however, 
that 
the quoted limiting fluxes are often not uniform over the entire areas).\\
The 
identification work and subsequent spectroscopy are very demanding
in terms of telescope time.
Typically, no more than $\sim 50-60\%$ of  the radio sources in sub-mJy 
samples have been identified on optical images, even though for the $\mu$Jy 
survey in the Hubble Deep Field 
an identification rate of about 80$\%$ has been reached (Richards et al. 
\cite{Richards99}).  
On the other hand, the typical fraction of spectra available is only 
$\sim 20\%$. The best studied sample is the Marano 
Field, where $\sim 45\%$ of the sources have spectral information (Gruppioni
et al. \cite{Gruppioni99a}). \\
To establish a firm point in the radio properties of galaxies in the local
(z $<$ 0.2) universe it is necessary to survey a large area in the sky down to 
 faint flux limits. Furthermore it is necessary to have 
in the same region a statistically significant  sample of galaxies 
with well studied optical properties (radial velocities, magnitudes etc.).
To alleviate the identification work, regions with 
deep photometry (possibly multicolor) already available provide a significant 
advantage.  
The region we have selected fulfills these requirements at least partially.\\
Vettolani et al. (\cite{Vettolani97})
made a deep redshift survey in two strips of $22^{\circ}\times 1^{\circ}$  and
$5^{\circ} \times 1^{\circ}$ near the SGP by studying 
photometrically and spectroscopically nearly all galaxies down to $b_J 
\sim$ 19.4. The survey, yielding 3342 redshifts (Vettolani et al. 
\cite{Vettolani98}), 
has a typical depth of $z=0.1$ with 10$\%$ of the objects at
$z>0.2$ and is $90\%$ complete.
In the same region lies the \emph{ESO Imaging Survey} (EIS, Nonino et al. 
\cite{Nonino99})
Patch A (3.2 sq. degr.), consisting of deep images in the I band out of which
a galaxy catalogue $95\%$ 
complete to $I=22.5$ has been extracted. Further V band 
images are available over $\sim$ 1.5 sq. degr.\\
We used the 6 km configuration of the ATCA
to make a 20 cm radio continuum mosaic of the region covered by
the  ESP galaxy redshift survey. 
The ATESP radio survey has uniform 
sensitivity ($1\sigma$ noise level $\sim $79 $\mu$Jy). \\

The present paper essentially deals with a description of the survey, 
the observations, the mosaic technique and the data reduction. It 
is organized as follows. In Sect.~\ref{sec-design} the survey design, 
in particular with
respect to the mosaic technique is explained.  
In Sects.~\ref{sec-obs} and \ref{sec-datared} we present in detail the 
calibration of our 20 cm 
observations and the data reduction. 
We discuss the problems encountered and the solutions adopted. 
Sect.~\ref{sec-mosanalysis} is dedicated to the analysis of the mosaics. 
A summary is given in Sect.~\ref{sec-concl}. 

\begin{table}[t]
\caption[]{1.4 GHz mJy and sub-mJy radio surveys. \label{rad-surv}}
\begin{flushleft}
\begin{tabular}{ l l c c}
\hline 
\hline
\multicolumn{1}{c}{Survey}
& \multicolumn{1}{c}{References}
& \multicolumn{1}{c}{Area}
& \multicolumn{1}{c}{$S_{lim}$}\\
  &  & sq. deg. & mJy \\
\hline \\
NVSS & Condon et al. \cite{Condon98}   & 3 $\times 10^4$ & 2.5 \\
ELAIS N b & Ciliegi et al. \cite{Ciliegi99} & 4.22 & 1.15\\
FIRST & White et al. \cite{White97}   & 1550 & 1.0\\
ELAIS S & Gruppioni et al. \cite{Gruppioni99b} & 4.0 & 0.4 \\
VLA-NEP & Kollgaard et al. \cite{Kollg94} &  29.3 & 0.3 \\
PDF & Hopkins et al. \cite{Hopkins98}  &  3.0 & 0.2\\
Marano Field & Gruppioni et al. \cite{Gruppioni97} & 0.36 & 0.2 \\
ELAIS N a & Ciliegi et al. \cite{Ciliegi99} & 0.12 & 0.135\\
LBDS & Windhorst et al. \cite{Windhorst84} & 5.5 & 0.1 - 0.2\\
Lockman Hole & de Ruiter et al. \cite{Ruiter97} & 0.35  &  0.12\\
Lynx 3A & Oort \cite{Oort87}  &  0.8 & 0.1  \\
0852+17 & Condon \& Mitchell \cite{CM84} & 0.32 & 0.08 \\
1300+30 & Mitchell \& Condon \cite{MC85} & 0.25 & 0.08\\ 
HDF & Richards \cite{Richards99a} & 0.3 & 0.04 \\
 & & & \\
ATESP & this paper & 25.9 & 0.47 \\
\hline
\hline 
\end{tabular}
\end{flushleft}
\end{table}

\section{Survey design}\label{sec-design}

The radio observations were carried out with two main goals in mind.
The first aim was to detect the ESP galaxies in order to derive the 
`local' ($z\sim 0.1$) 
bivariate luminosity function. We therefore tried to keep the
sensitivity as uniform as possible over the whole ESP area, while
at the same time reaching flux densities well below $\sim 1$ mJy (see 
Sect.~\ref{sec-freqres}). \\
The second aim was to have a complete catalogue of faint radio sources 
in order to study the sub-mJy population through a programme of optical 
identification of complete radio source samples extracted from the ATESP 
survey, exploiting the available data, i.e. deep CCD images.\\
As the survey is intended to achieve uniform sensitivity over a large area 
it is necessary to make use of the mosaicing technique. 

\subsection{Mosaicing Technique}\label{sec-mostech}
 
\begin{figure}[t]
\resizebox{\hsize}{!}{\includegraphics{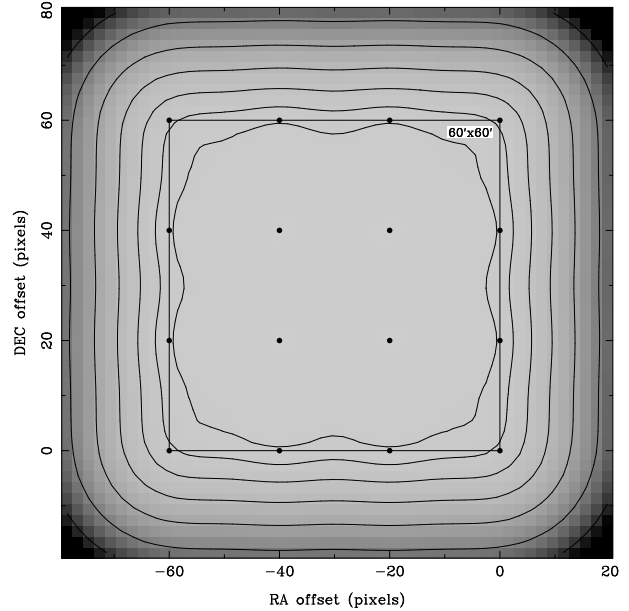}}
\caption{Noise variations expected over a 1~sq.degr. mosaic
(central box, 1 pixel $= 1\arcmin$) for a $20\arcmin$ spacing rectangular grid.
Filled circles indicate the pointing centres. At the center of the mosaic
$\sigma(I)/\sigma \sim 1$. Contours refer to 5, 10, 20, 40, 80, 160, 320\% 
noise increment from the center.} 
\label{fig-simrect}
\end{figure}

Mosaicing is the combination
of regularly spaced multiple pointings of a radio telescope which are then 
linearly combined to produce an image larger than the radio telescope's 
primary beam. The linear mosaicing consists of a weighted
average of the pixels in the individual pointings, with the 
weights determined by the primary beam response and the expected 
noise level (see equation in Sault \& Killeen \cite{SK95}). 
If the observing parameters and conditions are 
the same for every individual pointing, the expected noise variance in any
observed field can be assumed to be equal for every observing field and 
the intensity distribution in the mosaiced final image, $I(l,m)$, 
is modulated only by the primary beam response: 
\begin{equation}\label{eq-igrid2}
I(l,m) = \frac{ \sum_{i} P(l-l_i,m-m_i) I_i(l,m)}
         { \sum_{i} P^2(l-l_i,m-m_i)}
\end{equation}
where the summation is over the set of pointing centres $(l_i,m_i)$.
$I_i(l,m)$ is the image formed from the $i$-th pointing (not corrected for
the primary beam response) and $P(l,m)$ is the primary beam pattern. \\
In planning a mosaicing experiment, the main issue to be decided is
the pointing grid pattern, i.e. geometry and pointing spacings.
For a detection experiment on a
large area of sky (like the ATESP survey) the main requirement is   
uniform sensitivity over the entire region together with 
high observing efficiency. Such requirement can be satisfied by 
choosing opportunely the pointing grid pattern. \\
The mosaic noise standard deviation, $\sigma(I(l,m))$, can be obtained from 
the error propagation of Eq.~\ref{eq-igrid2} and the uniform sensitivity 
constraint is expressed by
\begin{equation}\label{eq-noisegrid2}
\frac{\sigma (I(l,m))}{\sigma} = \frac{ 1}{ \sqrt{\sum_{i} 
P_i^2(l-l_i,m-m_i)}} = constant
\end{equation}
for every position $(l,m)$. In other words, the mosaic sensitivity, 
expressed in terms of $\sigma$s ($\sigma$ is the noise expected in the 
individual pointings) is modulated by the squared sum of the primary beam 
response. \\
The ATCA primary beam pattern can be approximated by a circular Gaussian 
function (Wieringa \& Kesteven \cite{Wieringa92})
\begin{equation}
P_i (r) = \mathrm{e}^{-4\ln 2 \left( \frac{r}{\mbox{FWHP}} \right)^2}, 
\end{equation}
where $r=\sqrt{l^2+m^2}$ is the radial distance from the image
phase center and FWHP is the \emph{full width at half power} of the primary 
beam ($\sim$ 33 arcmin for ATCA observations at 1.4 GHz).
Since the square of a Gaussian is still a Gaussian with FWHP reduced by a
factor $\sqrt{2}$, the quadratic sum in Eq.~\ref{eq-noisegrid2}
can be written as a linear sum of Gaussians with 
FWHP$\arcmin = \mbox{FWHP}/\sqrt{2}$ (corresponding to $s\simeq
33\arcmin/\sqrt{2} = 23.3\arcmin$ for ATCA 1.4 GHz observations). \\
To make the final choice for the ATESP survey grid pattern, we have 
performed a series of simulations of the bidimensional quantity 
\begin{equation}\label{eq-sim}
\sigma(I)/\sigma = \frac{1}{\sqrt{\sum_i \mbox{e}^{-4\ln 2
\left(\frac{(s_i-r)}{\mbox{FWHP}\arcmin}\right)^2}}}
\end{equation}
where $s_i=(l_i,m_i)$, $r=\sqrt{(l-l_i)^2 + (m-m_i)^2}$ and 
FWHP$\arcmin=23.3\arcmin$, varying the pointing
spacings $s$ and the grid geometry (hexagonal and/or rectangular grids). \\
In general a very good compromise between
uniform sensitivity and observing efficiency is represented by a grid 
pattern with pointing spacings of the order of $s\simeq \mbox{FWHP}\arcmin$.
For our particular case, the best choice turned out to be a $20\arcmin$ 
spacing rectangular grid. The mosaic noise variations over a
reference area of 1~sq.~degr. for such a grid configuration are shown
in Fig.~\ref{fig-simrect}. 
As expected, in the region of interest (central box)
the noise is rather constant: variations are
$\leq 5\%$, except at the region borders ($\leq 10\%$). We point out that
hexagonal grids should be preferred when imaging wide 
areas of sky (like for the NVSS and FIRST radio surveys), but are not 
very efficient in the case of a narrow 1-degree wide strip of sky.

\noindent

\subsection{Observing Frequency, Resolution and Sensitivity}\label{sec-freqres}

At 20 cm (the longest
observing wavelength available at the ATCA) the field of view
is largest (gaussian primary beam FWHP$\simeq 33\arcmin$) and the system 
noise is lowest.
Thus, observing at 20 cm allows minimization of both the number of
fields (i.e. pointings) required to complete the survey and the observing 
time spent on each field.
The ATCA can observe at two frequencies simultaneously 
(for instance 20 and 13 cm). 
However, we decided to optimize sensitivity at the expense of spectral 
information, by setting both receivers in the 20 cm band and observing in 
continuum mode 
($2\times 128$ MHz bandwidth, each divided into $32\times 4$ MHz channels in 
order to reduce the bandwidth smearing effect). 
This choice was also influenced by another
consideration: since the field of view depends on the observing frequency,
the grid pattern could not be optimized to get uniform
sensitivity for both 13 and 20 cm bands simultaneously.\\
The observations were carried out at full resolution (ATCA 6 km
configuration), since the identification follow up benefits from high
spatial resolution, and the expected fraction of  
very extended sources (that could be resolved out and lost) is low at the 
ATESP resolution. Using the angular size 
distribution given by Windhorst et al. (1990) for radio sources, we estimate 
that $\ga 70\%$ of the mJy and sub--mJy sources would appear point--like
at the ATESP resolution, and $\la 5\%$ would have 
angular sizes twice the beam size or larger. \\
Since the ESP sample distance distribution peaks at $z\simeq 0.1$ and 
`normal' galaxies are typically low-power radio sources, deep radio
observations were needed to ensure detections of a statistically significant
number of ESP galaxies. We considered satisfactory a point source radio limit
of the order of $\sim 0.2$\,mJy ($3\sigma$), which corresponds to a 
detection threshold of $P \sim 3 \times 10^{21}$ W Hz$^{-1}$
at $z \sim 0.1$ (H$_{\rm o} =100$ km/s Mpc$^{-1}$). 
Furthermore  a large sample of sub-mJy radio sources can be constructed 
at a $6\sigma$ detection limit, corresponding to a flux limit of  
$\sim 0.5$ mJy.  

\section{Observations and calibration}\label{sec-obs}

To cover the two areas of $22\degr \times 1\degr$ (region A) and 
$5\degr \times 1\degr$ (region B) 
of the ESP survey with uniform sensitivity, $69\times 4$ and $15\times 4$
pointings at $20\arcmin$ spacing are needed respectively. An area of
1.3 sq.~degs. ($4\times 4$ pointings) of region A 
was not observed, because of the presence of the strong radio source 
PMNJ2326-4027,
which would have prevented us from reaching the deep noise level required. 
This reduced the total number of
fields to be observed to $65\times 4$ (region A) and $15\times 4$ (region B),
i.e. $320$ pointings in total.
The right ascension range actually covered by the ATESP survey is thus 
$23^{\mathrm{h}}
31^{\mathrm{m}}$ -- $01^{\mathrm{h}} 23^{\mathrm{m}}$ and 
$22^{\mathrm{h}} 32^{\mathrm{m}}$ to $22^{\mathrm{h}} 57^{\mathrm{m}}$
(J2000). \\
An observing time of 1.2 hours per pointing was needed to reach a
$3\sigma$ RMS noise level of $\simeq 0.2$ mJy (with $2\times 128$ MHz 
bandwidth).
To obtain good hour angle coverage we organized the snapshot observations as
follows: the 320 fields were divided in 16 sets of 20 fields each
($5\times 4$).
Every set was then observed for $2\times 12^{\mathrm{h}}$, that is, the 20 
fields were observed in sequence, for 1 minute each,
repeating this for 72 times and adding 3 minutes for
calibration every hour. \\
The observing campaign started in November 
1994 and was completed in January 1996. A log of the observations is 
given 
in Table~\ref{tab-log} (dates, arrays used, observing time and frequencies). 
We stress that, for our purposes, the use of different 6 km arrays is not 
relevant in any way. \\  
The two 128 MHz observing bands were set in the most interference-free
region of the 20 cm band (1.3--1.5 GHz). \\
The flux density calibration was performed through observations of the
source PKS B1934-638,
which is the standard primary calibrator for ATCA 
observations ($S=14.9$ Jy at $\nu = 1384$ MHz, Baars et al. \cite{Baarsetal77}
flux scale). 
The phase and gain calibration was based on observations of 
secondary calibrators, selected from the ATCA calibrator list. 
Every single $12^{\mathrm{h}}$ run and each of the two 
observing bands were calibrated separately following the standard procedures 
for ATCA observations. 

\begin{table}
\caption[]{Log of the observations. \label{tab-log}}
\begin{flushleft}
\begin{tabular}{ c r c c c}
\hline 
\hline
\multicolumn{1}{c}{Date}
& \multicolumn{1}{c}{$t_{\mathrm{obs}}$}
& \multicolumn{1}{c}{Array}
& \multicolumn{1}{c}{$\nu_1$}
& \multicolumn{1}{c}{$\nu_2$}\\
\multicolumn{1}{c}{}
& \multicolumn{1}{c}{$h$}
& \multicolumn{1}{c}{}
& \multicolumn{1}{c}{MHz}
& \multicolumn{1}{c}{MHz}\\
\hline
 & & & & \\
18/11/94--21/11/94 & $3\times 12$ & 6D & 1344 & 1452 \\
23/12/94--04/01/95 & $13\times 12$ & 6A & 1344 & 1452 \\
15/12/95--01/01/96 & $17\times 12$ & 6C & 1344 & 1448 \\
 & & & & \\
\hline
\hline 
\end{tabular}
\end{flushleft}
\end{table}

\section{Data reduction}\label{sec-datared}

For the data reduction we used the 
\emph{Australia Telescope National Facility} (ATNF) release of the 
\emph{Multichannel Image Reconstruction, Image Analysis and Display} (MIRIAD) 
software package (Sault et al. \cite{Setal95}). 
A number of steps in the reduction process are usually done 
interactively, specifically the removal of bad data (`flagging'), and, in a 
certain  measure, cleaning and self calibration. Due to the large amount of
data involved we found it more practical to develop a
semi-automated reduction pipeline.

\subsection{Flagging of bad data}\label{sec-flag}

\begin{table*}[t]
\caption[]{Main parameters for the 16 final mosaics. \label{tab-mospar}}
\begin{flushleft}
\begin{tabular}{ l c c c c r c c c }
\hline 
\hline
\multicolumn{1}{c}{Mosaic$^a$}
& \multicolumn{1}{c}{Fields}
&\multicolumn{2}{c}{Tangent Point$^b$}
& \multicolumn{2}{c}{Synthesized Beam$^c$}
&\multicolumn{1}{c}{$S_{\mathrm{min}}$}
& \multicolumn{1}{c}{$\sigma_{\mathrm{fit}}$}
&\multicolumn{1}{c}{$<\sigma>$}\\
\multicolumn{1}{c}{fld~$x$~to~$y$}
& \multicolumn{1}{c}{$n \times m$}
&\multicolumn{1}{c}{$\alpha_0$}
&\multicolumn{1}{c}{$\delta_0$}
& \multicolumn{1}{c}{$b_{\mathrm{min}}\times b_{\mathrm{maj}}$}
& \multicolumn{1}{c}{P.A.}
&\multicolumn{1}{c}{mJy}
& \multicolumn{1}{c}{$\mu$Jy}
&\multicolumn{1}{c}{$\mu$Jy}\\
\hline
 & & & & & & & & \\
fld01to06 & $6\times 4$ & 22 35 57.37 & -39 59 15.0 & $7.8\arcsec\times 12.9\arcsec$ &
$1\degr$  & $-0.49$ & 78.7 & $78.5\pm 3.0$\\
fld05to11 & $7\times 4$ & 22 44 57.60 & -39 59 15.0 & $7.8\arcsec\times 12.8\arcsec$ &
$3\degr$ & $-0.96$ & 77.8 & $78.5\pm 3.8$\\
fld10to15 & $6\times 4$ & 22 53 21.68 & -39 59 15.0 & $7.9\arcsec\times 13.0\arcsec$ &
$-1\degr$ & $-0.70$ & 88.1 & $84.1\pm 5.3$ \\
fld20to25 & $6\times 4$ & 23 34 56.09 & -39 58 48.0 & $8.5\arcsec\times 16.8\arcsec$ &
$1\degr$ & $-0.68$ & 83.0 & $80.3\pm 4.3$ \\
fld24to30 & $7\times 4$ & 23 43 38.26 & -39 58 48.0 & $9.5\arcsec\times 16.6\arcsec$ &
$-2\degr$ & $-0.50$ & 82.8 & $83.2\pm 6.4$\\
fld29to35 & $7\times 4$ & 23 52 20.43 & -39 58 48.0 & $8.9\arcsec\times 16.7\arcsec$ &
$2\degr$ & $-0.53$ & 79.2 & $76.2\pm 2.9$\\
fld34to40 & $7\times 4$ & 00 01 02.59 & -39 58 48.0 & $8.0\arcsec\times 14.5\arcsec$ &
$5\degr$ & $-0.48$ & 76.3 & $74.8\pm 4.0$\\
fld39to45 & $7\times 4$ & 00 09 44.76 & -39 58 48.0 & $8.6\arcsec\times 14.3\arcsec$ &
$-10\degr$ & $-0.61$ & 81.2 & $80.4\pm 3.4$\\
fld44to50 & $7\times 4$ & 00 18 26.93 & -39 58 48.0 & $7.4\arcsec\times 14.0\arcsec$ &
$7\degr$ & $-0.43$ & 78.0 & $76.4\pm 1.7$ \\
fld49to55 & $7\times 4$ & 00 27 09.10 & -39 58 48.0 & $8.0\arcsec\times 12.9\arcsec$ &
$10\degr$  & $-0.62$ & 78.6 & $77.1\pm 1.4$ \\
fld54to60 & $7\times 4$ & 00 35 51.26 & -39 58 48.0 & $8.2\arcsec\times 12.6\arcsec$ &
$4\degr$ & $-0.42$ & 77.3 & $75.8\pm 1.3$ \\
fld59to65 & $7\times 4$ & 00 44 33.42 & -39 58 48.0 & $7.7\arcsec\times 12.5\arcsec$ &
$4\degr$ & $-0.44$ & 79.4 & $77.8\pm 1.4$\\
fld64to70 & $7\times 4$ & 00 53 15.58 & -39 58 48.0 & $7.7\arcsec\times 12.6\arcsec$ &
$1\degr$ & $-0.45$ & 75.1 & $74.6\pm 2.2$ \\
fld69to75$^d$ & $7\times 4$ & 01 01 57.70 & -39 58 48.0 & $7.5\arcsec\times 12.8\arcsec$ &
$-2\degr$ & $-0.47$ & 81.9 & $79.3\pm 2.1$\\
fld74to80 & $7\times 4$ & 01 10 39.86 & -39 58 48.0 & $7.0\arcsec\times 15.0\arcsec$ &
$1\degr$ & $-0.61$ & 77.1 & $76.3\pm 3.8$\\
fld79to84 & $6\times 4$ & 01 19 22.03 & -39 58 48.0 & $6.8\arcsec\times 14.3\arcsec$ &
$7\degr$  & $-0.41$ & 68.9 & $67.6\pm 3.5$\\
 & & & & & & & & \\ 
\hline
\hline
\multicolumn{9}{l}{$^a$ $x$ and $y$ refer to the first and last field columns
composing the mosaic. $^b$ J2000 reference frame.} \\
\multicolumn{9}{l}{$^c$ P.A. is defined from North 
through East. $^d$ Reported values for $S_{\mathrm{min}}$ and noise refer 
to masked mosaic (see text).}
\end{tabular}
\end{flushleft}
\end{table*}

We made a modified version of the MIRIAD task TVFLAG (inserted in MIRIAD as 
TVCLIP),
which recursively flags visibilities with amplitudes exceeding a given 
threshold.
The threshold was set as a convenient multiple
of the average absolute deviation ($|\Delta S|$)
from the running median, evaluated separately for each baseline, each
channel and each integration cycle (10 s). \\
For the primary calibrator the automated flagging procedure was applied 
before the calibration. This was necessary to avoid the calibration being
affected by bad data.
For the secondary calibrators and the mosaic data the bandpass and 
instrumental polarization
calibration were applied before running TVCLIP. 
As we noticed that the shortest baselines introduced some low level, 
spatially correlated features
in the images, which could affect the zero level for faint sources, 
we decided to remove all baselines shorter than 
500 m from the data prior to the pipeline processing  
(rejection of $\sim 
10\%$ of the visibilities). As a consequence, the ATESP survey becomes 
progressively insensitive to sources larger than $30\arcsec$:  
assuming a Gaussian shape, only $50\%$ of the flux for a $30\arcsec$ large 
source would appear in the ATESP images. However,
the expected fraction of sources with angular sizes $\geq 30\arcsec$ is very
small: $\leq 2\%$ at fluxes $S\leq 1$ mJy according to the
Windhorst et al. (1990) angular size distribution. 

\subsection{Cleaning and Self-calibration}\label{sec-imag}

Since the primary beam response is frequency dependent, we did
not merge the data from the two observing bands before imaging and cleaning.
This results in a slightly poorer UV coverage but allows the cleaning 
process to succeed in subtracting correctly 100\% of the source flux, and
self-calibration to be more effective and reliable. \\
On the other hand, to improve UV coverage and sensitivity, for each field
we have merged the (calibrated) data coming from all the observing runs.\\
In contrast with the imaging of extended sources, 
joint deconvolution is not needed for a point source survey. It is also very
expensive computationally for high resolution images. We therefore
reduced every field separately, simplifying imaging considerably. \\
For each field a $2048\times 2048$ pixel image (total emission only)
was produced (pixel size = $2.5\arcsec$). The entire image was then 
cleaned in order to deconvolve all the sources in the field. To
improve
sensitivity we used natural weighting, which gave a
synthesized beam typically of 
the order of $8\arcsec \times 14\arcsec$.  \\
Each image went through different cleaning cycles. First, we
produced the list of the brightest components to use as model for
self-calibration. 
Phase only self-calibration was applied. Usually two iterations were
sufficient to remove phase error `stripes' and improve the image quality.\\
The self-calibrated visibilities were then used to produce a deeper
cleaned image.  
A serious problem arises if snapshot images are cleaned too 
deeply. 
Due to the incomplete UV coverage the number of CLEAN components can approach 
the number of
independent UV points. At this point the cleaning algorithm is not well 
constrained anymore
and can interprete noise (sidelobes, calibration errors, etc.)
as CLEAN components. This process can redistribute the noise into the
sources and, as a result, the process does not converge and,
in principle, the image can be cleaned to zero flux.
This produces many faint spurious sources, while the flux of real sources
is systematically underestimated. This effect has been
mentioned by Condon et al. (\cite{Condon98}) and White et al. (\cite{White97})
for deep snapshot observations with the VLA and is referred to as
`clean bias'. \\
From our tests we found that cleaning down to $3\sigma$, the
noise gets $\sim 20\%$ lower than theoretical. Down to $2\sigma$ it is a factor
of two lower and at $1\sigma$ can be 4--5 times lower. Thus we decided
to stop any cleaning process when the peak flux
residuals are of the order of 4--5 times the theoretical value
(setting a cut-off of 0.5 mJy) to minimize the clean
bias (a few percent effect expected, but see discussion in 
Sect.~\ref{sec-cleanbias}).\\
After this first phase of self-calibration and deep cleaning, we 
subtracted the sources from the visibility data and we repeated the
flagging procedure on the residual visibilities in order to reduce the
`birdies' effect mentioned in Sect.~\ref{sec-noise} below. \\
We then proceeded with another phase of cleaning, and produced a
half-resolution residual image covering 4 times the original area; only 
external parts of this
image were cleaned (not the inner quarter, corresponding to the original field,
which was already cleaned down to 0.5 mJy). 
This procedure allowed us to remove the sidelobes from 
more distant sources (belonging to adjacent fields).
These new components were subtracted from visibility data before
restoring the sources in the final cleaned image. \\
As a final step we checked for bright, extended sources in the field, which 
needed deeper cleaning. A small box containing such a source was cleaned, to 
a $2\sigma$ level.
In general, the application of all these cleaning steps produced good quality
single field images. 

\begin{figure*}
\vspace{23truecm}
\caption[]{One of the $6\times 4$ fields mosaiced images. The rectangular box 
indicates the region corresponding to the ESP redshift survey.
\label{fig-mos}}
\end{figure*}

\subsection{Mosaicing}\label{sec-mos}

The cleaned single field images were co-added in mosaics
in order to improve the signal to noise ratio and get uniform sensitivity.
Each set of $5\times 4$ fields observed in $2\times 12^{\mathrm{h}}$ observing
blocks produces a separate independent mosaic;
an overlap between adjacent mosaics was created by adding one (or two) 
extra column(s) of fields to the side(s) of each mosaic. \\
Before any mosaic is produced, every field was restored using the same
values for the beam parameters. The restoring parameters were chosen
as the average value of all the fields composing the mosaic at both
frequencies. \\
The final mosaics were obtained in two steps. First a single frequency
mosaiced image was produced for each of the two observing bands, then
the final mosaic was obtained by averaging (pixel by pixel)
the two initial mosaics in order to improve the sensitivity. \\
One of our final mosaics is shown in Fig.~\ref{fig-mos} 
as an example. The rectangular box indicates the region corresponding to the 
ESP redshift survey, where the radio survey was designed to give uniform noise.
Such a region covers an area of $1.67\degr \times 1\degr$ or 
$2\degr \times 1\degr$
for mosaics composed by $6\times 4$ or $7\times 4$ fields respectively (see 
Table~\ref{tab-mospar}).
Hereafter we will always refer to the central box only in our mosaic analysis.
All mosaics are available through the ATESP page at {\tt 
http://www.ira.bo.cnr.it}. 

\section{Mosaic analysis}\label{sec-mosanalysis}

Table~\ref{tab-mospar} summarizes the main parameters for the final 16 
mosaics: for each mosaic are listed  the number of fields composing it
(columns $\times$ rows), the tangent point (sky position used for geometry 
calculations) and the synthesized beam (size and position angle).
The spatial resolution can vary from mosaic to mosaic 
depending on the particular array (6A, 6C or 6D) used in the
observations. The mean value for the synthesized beam 
is $\sim 8\arcsec \times 14\arcsec$.

\begin{figure}
\resizebox{6cm}{!}{\includegraphics{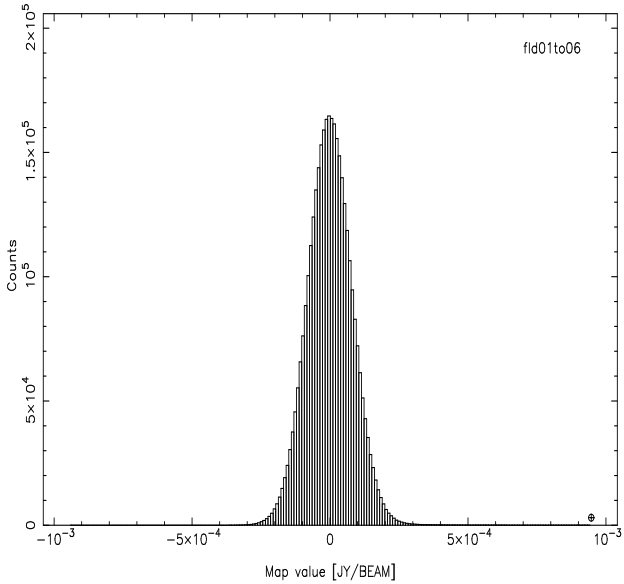}}
\caption[]{Histogram of the (residual) flux in one of the $6\times 4$ fields 
mosaiced images (fld01to06). As expected the flux is peaked at zero and 
the distribution is gaussian. 
\label{fig-noise}}
\end{figure}

\subsection{Noise}\label{sec-noise}

The last three columns of Table~\ref{tab-mospar} show the results of the 
noise analysis.
For each mosaic we report the minimum (negative) flux ($S_{\mathrm{min}}$) 
recorded on the image (typically $|S_{\mathrm{min}}|$ is of the order of $0.5$ mJy, 
corresponding to the value at which we have stopped the cleaning)
and the noise level. This has been evaluated either as the FWHM
of the gaussian fit to the flux distribution of the pixels (in the range
$\pm S_{\mathrm{min}}$),
 in order to check for correlated noise ($\sigma_{\mathrm{fit}}$), or
as the standard deviation of the average flux in several source-free
sub-regions of the mosaics,
in order to verify uniformity ($< \sigma >$). \\
As expected, the noise distribution is fairly uniform within 
each mosaic
and from mosaic to mosaic ($<10\%$ variations). Also, for each mosaic, the two
noise values are consistent, that is the noise can be considered gaussian (see also Fig.~\ref{fig-noise}). \\
On average the noise level is $\sim 79 $ $\mu$Jy. The typical detection
limits for the
ATESP survey are thus $\sim 0.24$ mJy at $3\sigma$ and $\sim 0.47$ mJy at
$6\sigma$.
Dynamic range problems can cause slightly higher noise levels of $\sim 100$
$\mu$Jy around strong sources ($S_{\mathrm{peak}}> 50-100$ mJy).
Such problems appear to be serious only in one mosaic (fld69to75): 
the region around the
bright radio source PMNJ0104-3950 
suffers from a very high noise level and
a number of spurious sources are present. This was due to strong phase 
instabilities during
the observations which could not be removed by self-calibration.
This region (of size $\sim 20\arcmin \times 25\arcmin$) was masked 
and therefore excluded from further analysis. Excluding this region,
the total unifom sensitivity area covered by the ATESP survey is $25.9$
sq. degr. or $7.9 \; 10^{-3}$ sr. 

\subsection{Artefacts}\label{sec-artefacts}

Another problem we faced was the possible presence of artefacts in
the images, like spurious sources (`ghosts') at a level of 0.1--0.5\% opposite 
 to bright sources with
respect to the phase center of the image and `holes' in the 
centre of the field. The first problem, caused by the Gibbs phenomenon, 
arises from the use of
an XF correlator and can be serious in high dynamic range continuum
observations (like ours). The second effect  
is a system error produced by the
harmonics of the 128 MHz sampler clock at 1408 MHz.
Both effects can be
completely removed as long as
the observing bands are centered appropriately (Killeen \cite{Killeen95};
Sault \cite{Sault95}). Unfortunately, at the time of our first two
observing runs these effects were not yet known. We therefore could apply the 
corrections only to the data taken in the last 
observing run. \\
We point out that the corrections, when applied, result in a larger bandwidth 
smearing effect, since only $13\times 8(10)$ MHz channels are used 
(instead of $32\times 4$ MHz).  \\
Wherever not corrected for, the `ghosts' problem is
unlikely to be serious in our case, since `ghosts' appear
in different places for each field and so they tend to average out 
when mosaicing the fields. 
Moreover, only radio sources
brighter than $\sim 100$ mJy can produce detectable `ghosts' in the final
mosaics and such bright sources are very few in the region surveyed ($\sim
30$) and therefore could be easily checked. No evident `ghosts' have been 
found.\\
We have also tried to reduce the sampler clock self-interference
effect as far as possible by flagging residual bad visibilities (that is
correlated noise) after a first
step of cleaning and self-calibration (see also Sect.~\ref{sec-imag}), 
but some 
of our fields still show it to a small extent. On the
other hand the area of sky affected by `holes' is of the order of a few
percents ($\la 3\%$) of the total region observed.

\subsection{Clean Bias}\label{sec-cleanbias}
  
\begin{table}
\caption[]{clean bias average corrections (see text).
\label{tab-cb}}
\begin{flushleft}
\begin{tabular}{ c c c c }
\hline 
\hline
\multicolumn{1}{c}{Mosaic}
&\multicolumn{1}{c}{cc's}
& \multicolumn{1}{c}{$a$}
&\multicolumn{1}{c}{$b$}\\
\hline
 & & &  \\
fld34to40 & $1616$ & $0.09 \pm 0.04$ & $0.85 \pm 0.08$ \\
fld44to50 & $2377$ & $0.13 \pm 0.03$ & $0.75 \pm 0.06$  \\
fld69to75 & $3119$ & $0.16 \pm 0.05$ & $0.67 \pm 0.09$  \\
 & & &    \\ 
\hline
\hline
\end{tabular}
\end{flushleft}
\end{table}

\begin{figure}
\resizebox{\hsize}{!}{\includegraphics{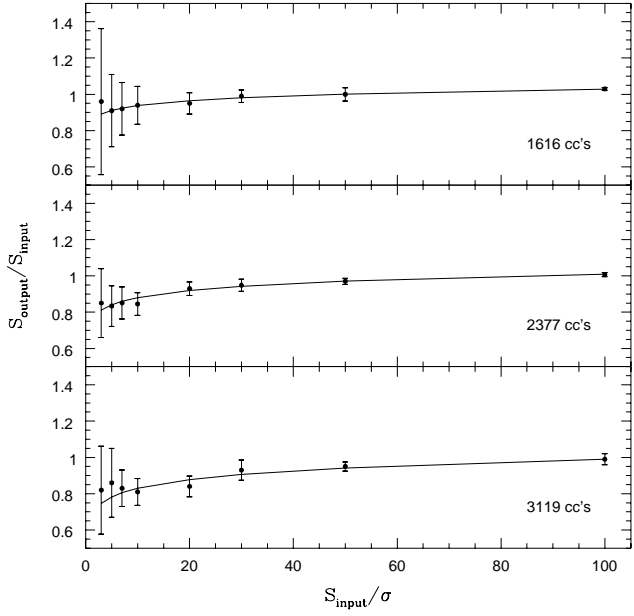}}
\caption[]{The source flux measured after the cleaning 
($S_{\mathrm{output}}$) normalized to the true source flux
($S_{\mathrm{input}}$) as a function
of the flux itself (expressed in terms of $\sigma$) is shown for three 
different cases: 1616 cc's mosaic (top panel), 2377 cc's mosaic 
(middle panel) and 3119 cc's mosaic (bottom panel). Also shown are the 
linear fits (see text).
\label{fig-cb1}}
\end{figure}

\begin{figure}
\resizebox{\hsize}{!}{\includegraphics{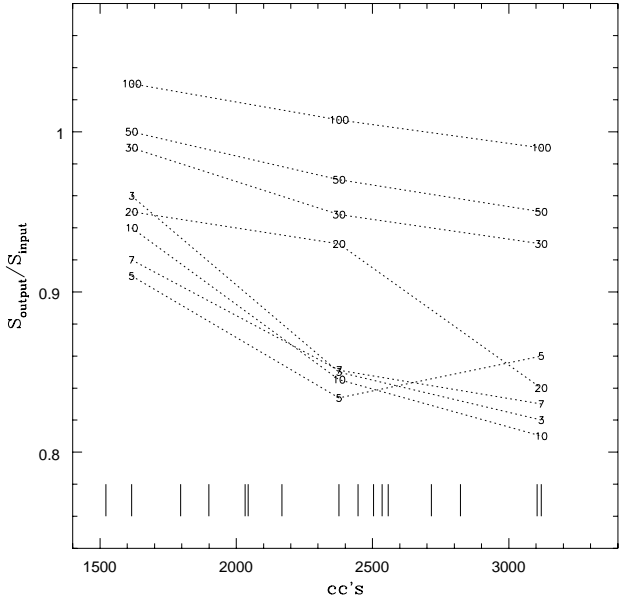}}
\caption[]{We present $S_{\mathrm{output}}/S_{\mathrm{input}}$
as a function of the average number of clean components. Each dotted line 
refers to a different source 
flux ($100\sigma$, $50\sigma$, $30\sigma$, etc.). Also shown is the average 
number of clean components for each of our 16 mosaics (full lines at the 
bottom of the figure).
\label{fig-cb2}}
\end{figure}

As already mentioned in Sect.~\ref{sec-imag}, 
when the UV coverage is incomplete,
the cleaning process can `steal' flux from real sources and redistribute it
on top of noise peaks producing spurious ones.  \\
To quantify the actual effect in our mosaics we performed a set of simulations
by injecting point sources in the survey UV data at random positions.  
Then the whole cleaning process was started. The number of sources 
injected in a mosaic ($\sim 500$) was chosen in order to reduce the 
statistical uncertainty without changing significantly the components/image 
statistics. The source fluxes cover the entire range of the survey: 250 
sources at 3$\sigma$, 100 at $5\sigma$, 50 at $7\sigma$, 50 at $10\sigma$, 
25 at $20\sigma$, 10 at $30\sigma$, 10 at $50\sigma$, 10 at $100\sigma$.  \\
Taking into account the time and frequency resolution and the baseline lenghts,
we get about 1500--2500 indepentent UV points for each field observed. This
means that with about 2000 \emph{independent} (not too close together or on 
top of each other) clean components we could clean the image to zero flux.  
The average number of (not independent) clean components per mosaic 
ranges between 1500 and 3200.  
Since we used a clean loop gain 
factor $g=0.1$, the number of independent clean
components per mosaic can be estimated as 1/10 of the numbers reported above. 
We thus expect a significant clean bias effect (10-20\%), larger for mosaics 
with a higher number of cleaning components.
We then decided to test three 
mosaics, with a low, an intermediate and a high average number of 
cleaning components (cc's) respectively:
fld$34$to$40$ (1616 cc's), fld$44$to$50$ (2377 
cc's) and fld$69$to$75$ (3119 cc's).\\
The results of the tests are presented in Fig.~\ref{fig-cb1} and 
Fig.~\ref{fig-cb2} and summarized in Table~\ref{tab-cb}.
Fig.~\ref{fig-cb1} shows, for each of the three mosaics,  
the average source flux measured after the cleaning 
($S_{\mathrm{output}}$) normalized to the true source flux 
($S_{\mathrm{input}}$) as a function
of the flux itself (expressed in terms of $\sigma$). In general the clean 
bias increases going to fainter fluxes and, as expected, depends on the 
number of cleaning components. In the best case (1616 cc's) we get $\leq 10\%$
flux underestimation for the faintest sources; in the worst case (3119 cc's)
the effect rises up to $\sim 20\%$. \\
The dependency of the clean bias on the number of cleaning components is more
clearly shown in Fig.~\ref{fig-cb2}.  
Here we present 
$S_{\mathrm{output}}/S_{\mathrm{input}}$
as a function of the average number of clean components for different source
fluxes ($100\sigma$, $50\sigma$, $30\sigma$, etc.).
Again, the clean bias affects more seriously the faintest sources. 
Moreover it is evident that we are not dealing with a linear effect: a 
sudden worsening appears at fluxes of the order of 10--20$\sigma$ and when 
the number of cleaning components exceeds $\sim 2000$.  \\
A first order fit of the clean bias effect for the three different mosaics
has been obtained by applying the least squares method to the function 
$S_{\mathrm{output}}/S_{\mathrm{input}}=a\log(S_{\mathrm{input}}/\sigma)+b$. 
The values obtained for the parameters $a$ 
and $b$ are listed in Table~\ref{tab-cb} and the curves are shown in 
Fig.~\ref{fig-cb1}.

\section{Summary}\label{sec-concl}

The ATESP survey at 1.4 GHz is based on
snapshot observations of 320 overlapping primary beam fields,
reduced separately and then combined together to
produce 16 big mosaiced images. 
The total area surveyed with uniform sensitivity ($1\sigma$ noise level 
$\sim 79 \mu$Jy)
covers 25.9 sq. degr.
The spatial resolution is typically $\sim 8\arcsec \times 14\arcsec$ providing
radio positions with internal accuracy of the order of 1 arcsec for  
$6\sigma$ radio sources. \\
We stress the importance of estimating the relevance and the behaviour of
the so called clean bias effect for any deep survey
obtained with snapshot observations. The source fluxes can be seriously 
affected by such problem and for reliable scientific analysis the effect
must be taken into account and corrected for.  \\

Future papers in this series will deal with:
\begin{enumerate}
\item the radio sources catalogue complete down to a limiting $6\sigma$ flux
density of $\sim 0.5$~mJy
\item the ATESP radio source counts 
\item the radio properties of the ESP galaxies
and the local bivariate luminosity function 
\item  the optical identifications and spectroscopy of the objects in the
EIS sub-region.
\end{enumerate}

\begin{acknowledgements}
IP would like to thank the ATNF and the ATCA for hospitality in Epping and 
Narrabri for long periods during 1994, 1995 and 1996. A special thank also to 
the ATNF staff, in particular to Neil Killeen and Bob Sault, for their 
valuable help in the development of the data reduction pipeline.
The authors acknowledge Roberto Fanti for reading and commenting on an 
earlier version of this manuscript. 
This project was undertaken under the CSIRO/CNR Collaboration programme.
The Australia Telescope is funded by the Commonwealth of Australia for 
operation as a National Facility managed by CSIRO.\\
\end{acknowledgements}

\end{document}